\documentstyle[twocolumn,epsfig]{mn}
\newif\ifAMStwofonts

\def\cm3{\rm cm^{-3}}

\def\msun{M$_\odot$ } 
 
\ifoldfss 
  \ifCUPmtlplainloaded \else 
    \NewTextAlphabet{textbfit} {cmbxti10} {} 
    \NewTextAlphabet{textbfss} {cmssbx10} {} 
    \NewMathAlphabet{mathbfit} {cmbxti10} {} 
    \NewMathAlphabet{mathbfss} {cmssbx10} {} 
  \fi 
  \ifAMStwofonts 
    \ifCUPmtlplainloaded \else 
      \NewSymbolFont{upmath} {eurm10} 
      \NewSymbolFont{AMSa} {msam10} 
      \NewMathSymbol{\upi}     {0}{upmath}{19} 
      \NewMathSymbol{\umu}     {0}{upmath}{16} 
      \NewMathSymbol{\upartial}{0}{upmath}{40} 
      \NewMathSymbol{\leqslant}{3}{AMSa}{36} 
      \NewMathSymbol{\geqslant}{3}{AMSa}{3E}

      \let\leq=\leqslant \let\le=\leqslant 
        
    \fi 
  \fi 
\fi 
 
\ifnfssone 
  \newmathalphabet{\mathit} 
  \addtoversion{normal}{\mathit}{cmr}{m}{it} 
  \addtoversion{bold}{\mathit}{cmr}{bx}{it} 
  \newmathalphabet{\mathbfit} 
  \addtoversion{normal}{\mathbfit}{cmr}{bx}{it} 
  \addtoversion{bold}{\mathbfit}{cmr}{bx}{it} 
  \newmathalphabet{\mathbfss} 
  \addtoversion{normal}{\mathbfss}{cmss}{bx}{n} 
  \addtoversion{bold}{\mathbfss}{cmss}{bx}{n} 
  \ifAMStwofonts 
    \ifCUPmtlplainloaded \else 
      \UseAMStwoboldmath 
      \makeatletter 
      \new@mathgroup\upmath@group 
      \define@mathgroup\mv@normal\upmath@group{eur}{m}{n} 
      \define@mathgroup\mv@bold\upmath@group{eur}{b}{n} 
      \edef\UPM{\hexnumber\upmath@group} 
      \new@mathgroup\amsa@group 
      \define@mathgroup\mv@normal\amsa@group{msa}{m}{n} 
      \define@mathgroup\mv@bold\amsa@group{msa}{m}{n} 
      \edef\AMSa{\hexnumber\amsa@group} 
      \makeatother 
      \mathchardef\upi="0\UPM19 
      \mathchardef\umu="0\UPM16 
      \mathchardef\upartial="0\UPM40 
      \mathchardef\leqslant="3\AMSa36 
      \mathchardef\geqslant="3\AMSa3E 

      \let\leq=\leqslant \let\le=\leqslant 

    \fi 
  \fi 
\fi 
 
\ifnfsstwo 
  \DeclareMathAlphabet{\mathbfit}{OT1}{cmr}{bx}{it} 
  \SetMathAlphabet\mathbfit{bold}{OT1}{cmr}{bx}{it} 
  \DeclareMathAlphabet{\mathbfss}{OT1}{cmss}{bx}{n} 
  \SetMathAlphabet\mathbfss{bold}{OT1}{cmss}{bx}{n} 
  \ifAMStwofonts 
    \ifCUPmtlplainloaded \else 
      \DeclareSymbolFont{UPM}{U}{eur}{m}{n} 
      \SetSymbolFont{UPM}{bold}{U}{eur}{b}{n} 
      \DeclareSymbolFont{AMSa}{U}{msa}{m}{n} 
      \DeclareMathSymbol{\upi}{0}{UPM}{"19} 
      \DeclareMathSymbol{\umu}{0}{UPM}{"16} 
      \DeclareMathSymbol{\upartial}{0}{UPM}{"40} 
      \DeclareMathSymbol{\leqslant}{3}{AMSa}{"36} 
      \DeclareMathSymbol{\geqslant}{3}{AMSa}{"3E} 

      \let\leq=\leqslant \let\le=\leqslant 

    \fi 
  \fi 
\fi 
 
\ifCUPmtlplainloaded \else 
  \ifAMStwofonts \else 
    \def\upi{\pi} 
    \def\umu{\mu} 
    \def\upartial{\partial} 
  \fi 
\fi 

\def\spose#1{\hbox to 0pt{#1\hss}}
\def\lta{\mathrel{\spose{\lower 3pt\hbox{$\mathchar"218$}}
     \raise 2.0pt\hbox{$\mathchar"13C$}}}
\def\gta{\mathrel{\spose{\lower 3pt\hbox{$\mathchar"218$}}
     \raise 2.0pt\hbox{$\mathchar"13E$}}}
 
\title{Three-dimensional simulations of the interstellar medium in dwarf 
galaxies - I. Ram pressure stripping}

\author[A. Marcolini, F. Brighenti and A. D'Ercole] 
       {A. Marcolini$^1$, F. Brighenti$^1$ and A. D'Ercole$^2$\\ $^1$Dipartimento di
       Astronomia, Universit\`a di Bologna, via Ranzani 1, 44127
       Bologna, Italy\\ $^2$Osservatorio Astronomico di Bologna, via
       Ranzani 1, 44127 Bologna, Italy}
 
\date{Accepted ..., Received ...; in original ...} 
 
\pagerange{\pageref{firstpage}--\pageref{lastpage}} 
\pubyear{2000} 
 
\begin{document} 
 
\maketitle 
 
\label{firstpage} 
 
\begin{abstract}
We present 3D hydrodynamic simulations of ram 
pressure stripping in dwarf galaxies.
Analogous studies on this subject usually deal with much higher
ram pressures, typical of galaxy clusters, or mild ram pressure due
to the gas halo of the massive galactic companions.
We extend over previous investigations by considering flattened,
rotating dwarf galaxies subject to ram pressures typical of
poor galaxy groups.

We study the ram pressure effects as
a function of several parameters such as galactic mass and velocity,
ambient gas density, and angle between the galactic plane and the
direction of motion. It turns out that this latter parameter plays a
role only when the gas pressure in the galactic centre is comparable
to the ram pressure. Despite the low values of the ram pressure, some
dwarf galaxies can be completely stripped after 1-2 hundred of million
years. This pose an interesting question on the aspect of the
descents and, more in general, on the morphological evolution of
dwarf galaxies. In cases in which the gas is not completely stripped, the propagation
of possible galactic wind may be influenced by the disturbed distribution
of the interstellar matter.

We also consider the modification of
the ISM surface density induced by the ram pessure and find that 
the resulting compression
may trigger star formation over long time spans.
\end{abstract}
 
 
\section{Introduction}

Galaxies are not closed boxes and their
evolution is closely connected with episodes of interstellar matter
(ISM) losses. These losses are due to internal mechanisms such as
galactic winds, or external mechanisms as the ram pressure stripping
exerted on a galaxy moving through the external
intracluster/intergalactic medium (ICM/IGM).

Since dwarf galaxies have small escape velocities, both galactic
winds and ram pressure stripping are expected to be more efficient
than for giant galaxies. Dwarf irregulars (dIrrs) often
have lower abundances than those predicted by the closed box chemical models
(cf. Matteucci \& Chiosi 1983).
This is
generally taken as evidence of metal enhanced outflow via galactic
winds, altough the
same effect can also be achieved by stripping of the gas (Skillman 2001).

The effect of the ram pressure on a dwarf galaxy evolution goes
beyond the mere reduction of the ISM mass. 
In fact, the ram pressure may influence substantially the dynamics
of galactic outflows, both in models in which the wind breaks out
of the galactic disk (e.g. De Young \& Heckman
1994, MacLow \& Ferrara 1999, D'Ercole \& Brighenti 1999, Strickland
\& Stevens 2000) and models in which the wind is suffocated by
an hypothesized extended gaseous halo and remains trapped in the
neighborood of the galaxy (Tenorio-Tagle 1996, and Silich \& 
Tenorio-Tagle 2001). Therefore the ram pressure may be an important
factor for the galactic
chemical enrichment. A detailed study of the ram pressure - 
galactic wind interaction will be presented elsewhere 
(Marcolini, Brighenti and D'Ercole, in preparation).
In the present paper we focus on the ISM removal by ram
pressure, which is an interesting phenomenon on its own. 

In fact, besides the interaction with the galactic wind, the ICM ram pressure
can be important also in triggering the starburst which gives rise
to the wind itself. Numerical simulations carried with the
smoothed-particle hydrodynamics (Abadi, Moore and
Bower 1999, Shulz \& Struck 2001) and with
$N$-body codes (Vollmer et al. 2001) show that, for small inclination
angles (nearly edge-on stripping), the ram pressure leads to a
temporary increase of the central gas surface density. 
This, in turn, may give rise to an episode of star formation.

Ram pressure stripping is also advocated to claim the dIrrs and
``transition-type dwarfs'' as progenitors of the dwarf spheroidal
(dSph) galaxies (Grebel, Gallagher and Harbeck 2003).
These latter galaxies experienced star formation over
extended time spans in their youths, but today they are free of detectable
ISM.
Being the galactic winds generally unable to remove a large fraction of the
galactic gas (Mac-Low \& Ferrara 1999; D'Ercole \& Brighenti 1999),
an external mechanism as ram pressure has been suggested
to strip the ISM. The
``transition-type dwarfs'' have mixed dIrr/dSph morphologies, low stellar
masses, low angular momentum, and HI content of at most a few $10^6$
$M_{\odot}$. These dwarfs would closely resemble dSphs if their gas were
removed, and are thus likely dSph progenitors.

Several papers investigated the gas dynamics of massive spiral
galaxies moving through the ICM of rich clusters (cf. Schulz \& Struck
2001 and references therein).  Few investigations discussed the effect
of stripping for dwarf galaxies. Mori \& Burkert (2000) and Murakami
\& Babul (1999) considered spherical dwarf galaxies moving through a
dense ICM with velocities $\sim 1000$ km s$^{-1}$, typical of clusters
of galaxies. Sofue (1994a,b) studied the moderate stripping on
spherical dwarf galaxies due to their motion through the gas halo of
larger companions. He adopted an $N$-body code and investigated the
different behaviour of HI and molecular gas, but some genuine hydrodynamic
phenomena such as shock waves and Kelvin-Helmholtz (K-H) instabilities are
inevitably ignored.

In this work we make use instead of pure hydrodynamical simulations
(cf. the discussion at the end of section 6), and study {\it disky} 
dwarf galaxies orbiting
inside small galaxy groups, where most of these galaxies reside
(Nolthenius 1993). Even dwarfs now in clusters have likely
spent a significant fraction of their life in loose groups,
before enter a cluster as the hierarchical
growth of the structures proceeded.

Velocity dispersions in groups are much lower
than in rich clusters, and the ram pressure stripping is much reduced.
However, even though the low ram pressure in groups is often unable
to completely remove the galactic gas, it may be still
effective in stripping the outer part of the ISM and affecting
the gas distribution inside the galaxy (Bureau and Carignan 2002,
Hidaka \& Sofue 2002). Moreover, ram pressure
due to gaseous halos of the Milky Way and M31 has been advocated by
Van den Bergh (1994) to explain the observed correlation between
stellar content and galactocentric distance of dwarf galaxies.

To investigate this important evolutionary phase in the life of dwarf
galaxies, we have undertaken a systematic study of the interaction
between IGM/ICM and the ISM. We run a number of 3D and 2D hydrodynamic
simulations in which we varied several parameters, as the galaxy mass,
the ram pressure strength, and the inclination between the galaxy and the
direction of its motion.

\section{the  model} 
\subsection{Potential well} 
\noindent 
The gravitational potential of our models is due to the contribution of 
two mass distributions: a spherical quasi-isothermal dark matter halo 
and a stellar thin disk. The dark matter halo density is given by: 
\begin{equation}
\rho_{\rm h}(r)=\frac{\rho_{\rm 0h}}{[1+(\frac{r}{r_{\rm h}})^2]},
\end{equation}
where $\rho_{\rm 0h}$ is the central density and $r_{\rm h}$ is the 
core radius. The distance to the centre is given by $r=\sqrt{R^2+z^2}$, 
where $R$ is the radius on the equatorial plane ($z=0$).  The dark 
halo is truncated at a radius $r_{\rm tr}$, and its
mass as a function of $x=r/r_{\rm h}$ is then:
\begin{equation}
M_{\rm h}(r)=
4 \pi \rho_{\rm 0h} r_{\rm h}^3(x- \arctan x).
\end{equation}
With these assumptions the 
gravitational potential of the halo is given by:
\begin{equation}
\Phi_{\rm h}(r) = \left\{ \begin{array}{ll} 4 \pi G \rho_{\rm 0h}
r_{\rm h}^2 \{ -1 + \frac{\arctan x}{x}  + \\
\frac{1}{2} [\log (1+x^2)-\log(1+\psi^2)]
\}, & r \leq r_{\rm tr} \\ \\
- \frac {G M_{\rm h,tot}}{r}, & r > r_{\rm tr} \end{array} \right.
\end{equation}
where $\psi=r_{\rm tr}/r_{\rm h}$ 
and $M_{\rm h,tot}=M_{\rm h}(r_{\rm tr})$.

We assume that the stars are distributed in an infinitesimally thin 
Kuzmin's disk with surface density (cf. Binney \& Tramaine 1987):
\begin{equation}
\Sigma_*(R)= \frac {a M_{*} }{2 \pi (R^2+a^2)^{3/2}},
\end{equation}
where $M_{*}$ is the total mass 
of stars and $a$ is a radial scale-length. The           
stellar potential generated by this mass distribution is:
\begin{equation}
\Phi_* (R,z)=-\frac{ G M_{*} }{ \sqrt {R^2 + (a + |z|)^2}}.
\end{equation}
We consider three different galactic models differing mainly for their 
masses and hereafter called small (SM), intermediate (MD) and large (LG) 
(see Table 1).

\subsection{Gas distribution} 
\noindent 
The ISM is assumed to be single-phase, isothermal and in
rotational 
equilibrium with the potential $\Phi$ generated by the dark matter 
halo plus the stellar thin disk.

To obtain a realistic profile of the rotation curve and the surface 
density we procede as follow (cfr. D'Ercole \& Brighenti 1999).
First we assume a gas distribution in the equatorial plane ($z=0$) 
of the form:
\begin{equation}
\rho(R,0)= \frac{\rho_0}{ (1+(\frac{R}{R_{\rm c}})^2)^2}.
\end{equation}
Then, in order to build a rotating ISM configuration in equilibrium
with the given potential $\Phi$, we solve the steady state momentum equation:
\begin{equation}
(v \cdot \nabla) v = -\frac{1}{\rho} \nabla(P) - \nabla{\Phi}.
\end{equation}
\noindent
The rotational velocity in the equatorial plane is given by the
equilibrium condition:
\begin{equation}
v_\phi^2(R,0)=v_{\rm cir}^2-\frac{R}{\rho}\left|\frac{dP}{dR}\right|_{z=0},
\end{equation}
\noindent
where $v_{\rm cir}=\sqrt{R\partial \Phi/\partial R}$ is the circular 
velocity and $P$ 
the thermal gas pressure.
The gas distribution
out of the plane is obtained integrating the $z$ component of the
hydrostatic equilibrium equation for any $R$. The resulting gaseous disk
shows a pronounced flare (see upper panels $a$ in
Fig. 4, discussed below), a consequence of 
the assumption that $v_\Phi$ does not depend on $z$. This assumption is
necessary to have an isothermal gas in rotational equilibrium
(Tassoul 1978). We stress that the basic results of our simulations
do not depend on the presence of this (somewhat unrealistic) flare.
With an appropriate choice of the parameters $\rho_0$
and $R_{\rm c}$ (Table 2) we obtain realistic radial profiles of the
column densities and circular velocities (Swaters 1999) 
for our three models, as shown in Fig. 1 and 2.

In the following, we shall investigate the variation of the mass of gas,
induced by the ram pressure stripping,
in two regions: the {\it galactic region} and the {\it central region}.
The first is intended to contain the ``total'' amount
of ISM $M_{\rm gal}$ associated to the galaxy. 
It is defined as a cylindrical volume
within $z<|z_{\rm gal}|$ 
and $R<R_{\rm gal}$, where the values of $|z_{\rm gal}|$ and $R_{\rm gal}$
for the various models are
given in Table 2.
The second, smaller region roughly represents the volume of the stellar disk
of the galaxy and provides a measurement of the central content of ISM
$M_{\rm centr}$.
It is defined by 
$z<|z_{\rm centr}|$ 
and $R<R_{\rm centr}$, where $|z_{\rm centr}|$ and $R_{\rm centr}$ are again given
in Table 2.
We believe that the analysis of gas removal in these two regions
is more meaningful and instructive that the simple computation of the
ISM in the computational grid, which extends to a very large distance from the
galactic centre (section 4).
The initial amount of gas in these two regions is given in Table 2.

Finally, we considered also a non rotating galaxy (model NR) 
having exactly the
same potential well of the MD model, but in which the ISM does not rotate.
This model was run to consider the effect of the presence of a spheroidal
gaseous
halo around the galaxy. Such an halo is important also in view of its
influence on the evolution of the possible galactic wind.
In order to obtain an halo extension
comparable to the galactic size, we adopted an ISM temperature which is
somewhat larger than that of the ISM in the MD model, as shown in Table 2.
The gas distribution is simply specified by the 
(isothermal) hydrostatic equilibrium
condition, with the central density given in Table 2.

\subsection{The ram pressure}
\noindent
We are interested in ram pressure values characteristic for poor galaxy 
groups.
Group velocity dispersions of few hundred km s$^{-1}$ are typical
(Zabludoff \& Mulchaey 1998). Characteristic IGM densities are 
$n \sim 10^{-4} - 10^{-3}$ cm$^{-3}$ (e.g. Mushotzky et al. 2003). 
We thus considered
for each of our models two values for the ram pressure 
$P_{\rm ram} = \rho_{\rm IGM} v_{\rm IGM}^2$ within the range of typical
values: $P_{\rm LOW} = 8 \times 10^{-14}$ dyn cm$^{-2}$ and
$P_{\rm HIGH} = 3.2 \times 10^{-12}$ dyn cm$^{-2}$. The first case is realised
with the density-velocity combination 
($\rho_{\rm IGM},v_{\rm IGM})=(2 \times 10^{-28} \rm{g} \, \rm{cm}^{-3}, 
200 \;\rm{km} \, \rm{s}^{-1}$), while the latter with
($\rho_{\rm IGM},v_{\rm IGM})=(2 \times 10^{-27} \rm{g} \, \rm{cm}^{-3},
400 \;\rm{km} \, \rm{s}^{-1}$).
The IGM parameters are summarized in Table 3.
The temperature of the IGM is $T_{\rm IGM}=10^6$ K for both cases.
This temperature is somewhat lower than that 
predicted by the observed $\sigma - T$ relation for galaxy groups
(Helsdon \& Ponman 2000), assuming $\sigma \sim v_{\rm IGM}/\sqrt{3}$.
The choice of
a low $T_{\rm IGM}$ has been guided by the desire to have galaxies moving
supersonically through the IGM. 
In this regime, any numerical perturbation
does not propagate upstream
and does not interfere with the ``inflow'' boundary condition there.

\begin{table*}
\centering
\begin{minipage}{110mm}
\caption{Galaxy parameters}
\begin{tabular} {|c|c|c|c|c|c|c|c|c|}
\hline
Model  & $\rho_{\rm 0h}$ & $r_{\rm h}$ & $r_{\rm tr}$ & $a$ & $M_*$ & $M_{\rm {h,tot}}$ \\
       & ($10^{-24}$ g cm$^{-3}$) & (kpc) & (kpc) & (kpc) & $(10^8\,M_{\odot})$ &$(10^9\,M_{\odot})$ \\
\hline
SM     &   1.4       & 0.45  & 15     & 0.9   &  0.6  & 0.76  & \\
MD     &   1.4       & 1.00  & 30     & 1.8   &  6.0  & 7.4  & \\
LG     &   1.4       & 2.30  & 60     & 3.6   & 60.0  & 77.2  & \\
NR     &   1.4       & 1.00  & 30     & 1.8   & 6.0   & 7.4  & \\
\hline
\end{tabular}
\end{minipage}
\end{table*}

\begin{table*}
\centering
\begin{minipage}{150mm}
\caption{ISM parameters}
\begin{tabular} {|c|c|c|c|c|c|c|c|c|c|}
\hline
Model& $\rho_0 $ & $ R_{\rm{c}}$ & $ T $ & $M_{\rm{centr}}$ $^{(a)}$ & 
     $M_{\rm{gal}}$ $^{(b)}$  & $R_{\rm{centr}}$ &  $|z_{\rm{centr}}|$  
     & $R_{\rm gal}$  &  $|z_{\rm gal}|$  \\ 
     & ($10^{-24}$ g cm$^{-3}$) & (kpc) & (10$^3$ K) &$(10^7\,M_{\odot})$ 
     &$(10^8\,M_{\odot})$ & (kpc) & (kpc) & (kpc) & (kpc) \\
\hline
SM   & 1.0 &  1.2 & 2.3 & 0.9 & 0.5 & 1.0 & 0.5 & 4.0 & 1.0 \\
MD   & 2.0 &  2.4 & 4.5 & 6.6 & 4.1 & 2.0 & 1.0 & 8.0 & 2.0 \\
LG   & 4.0 &  4.8 & 9.0 & 42.2& 33.1 & 4.0 & 2.0 & 16.0 & 4.0 \\
NR   & 2.0 &   -$^{(c)}$  & 10.0 & 1.3& 0.17 & 1.0$^{(d)}$ & 1.0$^{(d)}$ & 
4.0$^{(d)}$ & 4.0$^{(d)}$ \\
\hline
\end{tabular}
\par\noindent
$^{(a)}$ initial mass content of the central region defined as a cylinder with 
$R<R_{\rm centr}$ and $z<|z_{\rm centr}|$.\\
$^{(b)}$ initial mass content in the galactic region defined as a 
cylinder with $R<R_{\rm gal}$ and $z<|z_{\rm gal}|$.\\
$^{(c)}$ no value for $R_{\rm c}$ is specified for the static ISM of 
this model\\
$^{(d)}$ for the NR model the central and the galactic regions are spheres
with the specified radii.
\end{minipage}
\end{table*}

\begin{figure}
\begin{center}
\psfig{figure=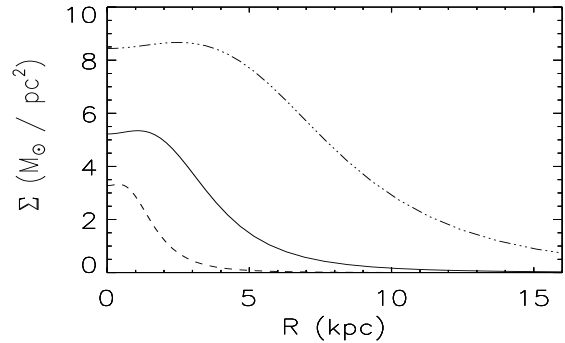,width=8cm,height=5cm}
\end{center}
\caption{Face-on column density profiles
of the initial ISM for the three galaxy 
models: SM (dashed line), MD (solid line) and LG (dashed dotted dotted line).}
\end{figure}

\begin{figure}
\begin{center}
\psfig{figure=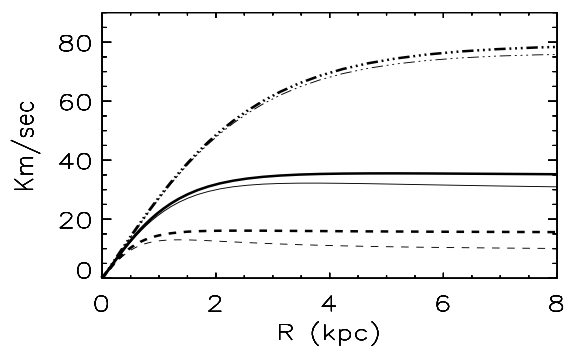,width=8cm,height=5cm}
\end{center}
\caption{Radial profiles of the circular $v_{\rm circ}$ (thick lines) and 
rotational $v_{\phi}$ velocities for the three galaxy models:
SM (dashed lines), MD (solid lines) and LG (dashed dotted dotted lines).}
\end{figure}

\section{Analytic considerations}

\noindent It is interesting
to make some simple prediction about the effectiveness of the ram pressure
stripping for both face-on and edge-on models.
As discussed by Mori \& Burket (2000), the stripping process may
be broadly classified into two regimes: the instantaneous
stripping and the continuous stripping assisted by
Kelvin-Helmholtz (K-H) instabilities.

Instantaneous stripping occurs if the ram pressure is larger
than the restoring gravitational force per unit area. In the case of
face-on models an analytical estimate of the radius $R_{\rm str}$ beyond
which the gas will be stripped
is obtained solving the equation (Gunn \& Gott 1972):
\begin{equation}
\frac{\partial \Phi}{\partial z}(R_{\rm str},z)\sigma_{\rm ISM}(R_{\rm str})=
P_{\rm ram},
\end{equation}
where the left-hand side is the restoring force for unit surface.
The value of the $z$ component of the gravitational acceleration
at any $R$ is chosen to be
the maximum along the $z$ direction, and $\sigma_{\rm ISM}$ is the ISM
surface density.
In Fig. 3 the restoring force is shown by the
continuous curves for the SM, MD and LG galactic models (from the
bottom to the top, respectively).  The two horizontal lines represent
the two values of the ram pressure considered.  The intersections of
these straight lines with the curves individuate $R_{\rm str}$.
If the ram pressure value lies above
the maximum of a curve, the galaxy
is completely stripped. This is similar to say that
$P_{\rm
ram}$ exceeds the thermal pressure $P_0$
at the centre of the gravitational potential
well.

As pointed out  by several authors (e.g. Schulz \&  Struck 2001, Vollmer et
al.  2001), the differences of the ram pressure effects are not simply
explained  by the projection  cosin law  in cases  where the  IGM wind
impacts the galactic disk at  a significant angle $\theta$ to the disk
symmetry  axis.  In the  case of  edge-on models,  we thus  followed a
different approach and considered the continuous stripping to obtain a
condition  similar to  equation (9). The most effective K-H modes
for stripping are those with wavelengths 
comparable to
the  galactic  radius  $R$  (Nulsen  1982).   These  modes,
however, could be suppressed by  the gravity
$g$ if  this is larger than a critical  value (Murray  et al.
1993):
\begin{equation}
g_{\rm cr}=\frac  {2 \pi v_{\rm IGM}^2} {\alpha R},
\end{equation}
where $\alpha$ is the ratio between the ISM density and the IGM density at
distance $R$. Equation (10) can be written as:
\begin{equation}
\frac {\rho(R_{\rm str}) g(R_{\rm str},0) R_{\rm str}}{2 \pi}=P_{\rm ram},
\end{equation}
and we can individuate the radius $R_{\rm str}$, at which the ram pressure
become less effective in removing the ISM, as the ``final'' radius
of the galaxy.
The dashed curves in Fig. 3 represent the left-hand side of the above
equation for the SM, MD and LG galactic models (from the bottom to the top,
respectively).
The outer intersection of these curves with the horizontal lines gives
the stable radius $R_{\rm str}$ for the galaxy .

Again, if a
particular value of the ram pressure lies above the maximum of these
curves, no stable radius can be found and eventually the galaxy will be
completely stripped after a characteristic time
\begin{equation}
\tau_{\rm str}\equiv M/\dot M \sim {R_{\rm gal} \bar \alpha  
\over v_{\rm IGM}},
\end{equation}
\noindent
where $\dot M$ is the (maximum) rate of mass loss, derived by momentum
conservation considerations (Nulsen 1982),
$\bar \alpha= \bar \rho /\rho_{\rm ICM}$, and $\bar \rho$ 
is the ISM mean density. Note that, contrary to what
reported by some author, this time scale (a lower limit) does not 
coincide with the
characteristic growth scale $\tau_{\rm KH}$ of the K-H instability, but is
longer by a factor $\bar \alpha^{1/2}$.

The couples of continuous (face-on galaxies) and dashed 
(edge-on galaxies) curves referring to each galaxy
model have similar maxima; this would indicate that the orientation
of the galaxies 
is rather uninfluential for the complete stripping of
the galaxy. However, our numerical simulations show instead
that the orientation strongly influences the effectiveness of the
stripping when $P_{\rm ram} \sim P_0$ (see section 5). 
Symbols in Fig. 3 represent the values of $R_{\rm
str}$ obtained by our numerical simulations discussed below.

We point out that the Reynold number $Re=2.8(r/\lambda_{\rm IGM}){\cal
M_{\rm IGM}}$ (Batchelor 1967) for the IGM is rather high. In fact,
for $T_{\rm IGM}=10^6$ K, Mach number ${\cal M_{\rm IGM}}\sim 1$
and $\rho_{\rm IGM} = 2 \times 10^{-27}\; (2 \times 10^{-28})$ g cm$^{-3}$,
the mean free path is $\lambda_{\rm IGM}\sim 3\times 10^{3}T^2n_{\rm
IGM}^{-1} \approx 1 \; (10)$ pc. With a
length scale $r=10$ kpc we get $Re\sim
10^4$ ($10^3$). These values are much larger than the critical
value $\sim 30$ (Nulsen 1982) at which the transition from laminar
to turbulent flows occurs. By contrast, the effective Reynold number
in our simulations is (cf. McKee 1988)
$Re\approx rv/c\Delta \sim 60$ (with a zone size
$\Delta=0.5$ kpc at $r=10$ kpc). Thus, our simulated 
flows are only marginally turbulent, and the
effectiveness of the K-H instability to drive the stripping could be
questioned. However, as shown by Nulsen (1982), mass stripping rate
for marginally laminar flows is comparable, within a factor of two,
with that occurring in fully turbulent flows.
We thus conclude that the 
stripping rates we obtain are still realistic.

\begin{table}
\centering
\begin{minipage}{65mm}
\caption{IGM parameters}
\begin{tabular} {|l|c|c|c|c|}
\hline
         &                      &         &             \\ 
Model    & $  \rho_{\rm IGM}$   &$v_{\rm IGM}$&  $T_{\rm IGM}$  \\
         &($10^{-24}$ g cm$^{-3}$)&  (km s$^{-1}$)  & (K)        \\
\hline
LO       & $2 \times 10^{-4} $  &    200  &   $ 10^6 $  \\
HI       & $2 \times 10^{-3} $  &    400  &   $ 10^6 $  \\
\hline
\end{tabular}
\end{minipage}
\end{table}

\section{The numerical method}

To run the set of simulations we used two hydrocodes: the 3D BOH
(BOlogna Hydrodynamics) code and ZEUS-2D (Stone \& Norman 1992).
The 3D BOH code uses an Eulerian, second order upwind scheme
(Bedogni \& D'Ercole 1986), in which the consistent
advection (Norman, Wilson \& Barton 1980) is implemented to reduce
numerical diffusion. The 3D version of the code has been tested
against a set of hydrodynamic problems with very satisfactory
results (Marcolini 2002).

For the models with edge-on or $45^\circ$ stripping we used the BOH code
(in Cartesian coordinates),
while for most face-on stripping models we used ZEUS-2D (in cylindrical
coordinates).
As a test, we run few face-on stripping models 
with both codes. We found that
the differences between the two simulations are negligible.

As most previous investigations (e.g. Mori \& Burkert 2000;
Quilis, Moore \& Bower 2000), we neglect radiative cooling.
In order to check whether this assumption affects our results,
we run a model including radiative losses (see section 5.1.2).
No appreciable differences have been found and we conclude
that neglecting radiative cooling is a safe assumption.

The 3D Cartesian grid adopts a non-uniform grid spacing.
The central zones are $\Delta x=\Delta y= \Delta z = 20$ pc wide,
and the zone size increases geometrically, the size ratio between 
adjacent zones being 1.1.
The $z=0$ plane
coincides with the galactic plane, the $z$-axis is the rotation axis. 
Each axis 
extends from $-30$ to 30 kpc and contains 106 zones
(for the edge-on models we considered only the volume
$0\le z\le 30$ kpc, with the $z$-axis subdivided in 54 zones). 

Given the symmetry of the simulation for edge-on models, we use
reflecting boundary conditions on the equatorial plane ($z=0$), inflow
conditions on the boundary plane from which the IGM enters the grid
($x=-30$ kpc plane), and outflow conditions on the
remaining planes.  For $45^{\circ}$ models, the inflow conditions are
enforced both at the $x=-30$ kpc plane and $z=30$ kpc plane,
while outflow conditions hold on the remaining boundary surfaces of the
grid.  For 2D face-on cylindrical symmetric models, we have reflecting
conditions on the $z$ axis, inflow conditions on the boundary plane
$z=30$ kpc, while outflow conditions are applied at the remaining
boundaries.

\section{Results}

We run 18 simulations for the rotating galaxies, varying the galactic
mass, the ram pressure strength, and the inclination angle $\theta$
between the galactic plane and galactic velocity.
We identify a particular model with the notation XX-YY-ZZ, where
XX individuates the galaxy size (SM=small, MD=medium, LG=large,
cf. tables 1 and 2); YY expresses the angle $\theta$, and takes the values
YY=00 for edge-on models, YY=45 for $\theta=45^\circ$, and YY=90
for face-on models.
Finally, ZZ represents the value of the ram pressure; ZZ=LO for the 
weak ram pressure, ZZ=HI for the high one (see table 3).

We also run two models for a non-rotating galaxy, one for
each value of the ram pressure (models NR-LO and NR-HI). In these
simulations the ram pressure hits the galaxy face-on.

Below we describe in some detail the gasdynamics of the models,
paying a special attention to the
the representative galaxy ``MD'', our reference model.

\subsection{The reference model}

\subsubsection{Model MD-00-LO}

Here we describe in some detail the hydrodynamic evolution of the
intermediate galaxy model moving edge-on and undergoing the action of the
lower ram-pressure. In the upper panels in Fig. 4 (upper strip) the density
distribution on the $y=0$ plane is shown at four different times. The
lower panels show the density distribution on the equatorial
plane at the same times. Tha IGM enters the grid from the left
boundary. The panels at $t=0$ represent the initial
conditions.
After $t=250$ Myr a rather complex structure is formed. The
interaction of the IGM with the ISM gives rise to two shocks: one
reflected into the IGM and forming the bow shock, and another
transmitted forward through the ISM. The bow shock is rather weak
given the low Mach number of the galactic motion (${\cal
M}=1.3$). The transmitted shock, initially with a similar Mach
number, propagates in a denser medium and is better
discernible in the lower panel $b$.  At later times, the lower
branch of the transmitted shock wraps quite easily around the core of
the galaxy on the equatorial plane, because of the anticlockwise
rotation of the ISM. Instead, the upper branch remains
nearly stationary (lower panel $c$) because is contrasted by the ISM
motion.  Later on (lower panel $d$) an asymmetric tail forms downstream.
The gas distribution on the disk assumes an elliptical
shape because of the compression due to the incoming IGM.  The
same effect is also apparent in the upper panel $d$ where the forward lobe
of the ISM is compressed and expands vertically, partially
``shielding" the rest of the galaxy by the interaction of the IGM.

\begin{figure}
\begin{center}
\psfig{figure=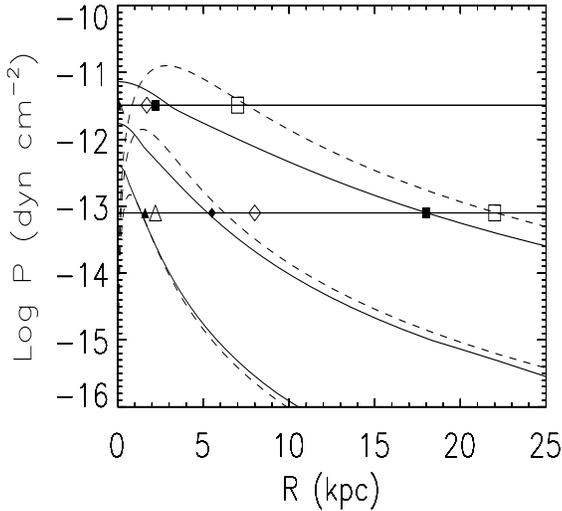,width=8cm,height=8cm}
\end{center}
\caption{The solid curves represent the radial profile of the maximum value of the $z-$component 
of the restoring force for our three galactic models SM, MD and LG (from the
bottom to the top, respectively) [equation (9)]. Analogously, 
the dashed curves represent the gravitational stabilising force of the 
K-H instabilities [equation (11)].
The horizontal lines represent the lower (LO) and higher (HI) 
ram pressure values in our models of IGM.
The intersections between the solid (dashed) curves and the horizontal lines
individuate the final radii in the case of face-on (edge-on) stripping. 
Symbols represent the radii found in our simulations: triangles,
diamonds and squares refer to SM, MD and LG models, respectively.
Filled symbols refer to face-on models, while empty symbols refer to
edge-on models.}
\end{figure}

In the present model both the instantaneous and the continuous
stripping are relatively ineffective in removing the ISM.
Figure 5 shows the ISM mass content in both the central and galactic regions;
in order to avoid the possible contribution of the IGM, only gas with
temperature $T<3\times 10^5$ K is considered in these plots. 
This figure shows that about $90 \%$ of the gas is retained
in the galactic region, while the central region is essentially unaffected.
The central portion of the
transmitted shock moving along the $x$ axis has a pressure which is of
the order of the ram-pressure, much lower than the ISM pressure $P_0$
in the central region. As the transmitted shock approaches this
region, it becomes weaker and weaker and eventually stalls as a sonic
perturbation.  The instantaneous stripping is thus effective only at the
very outskirts of the galaxy. 

The continuous stripping in this model is also rather ineffective,
being $\tau_{\rm str}>> 1$ Gyr. 
As pointed out in section 3, in our simulations the continuous
stripping occurs essentially via numerical viscosity, while for
real galaxies the turbulent stripping is more likely. However,
the viscous stripping rate is expected to be comparable with the turbulent
stripping (Nulsen 1982). In this latter case the K-H instabilities
would be strongly suppressed by the gravity, being $g\gta g_{\rm cr}$.

\subsubsection{Models MD-45-LO and MD-90-LO}

The maps of the gas density for model MD-45-LO are shown in the
middle strip of Fig. 4. 
In the upper panels the IGM enters the grid from the left
and top boundaries, and crosses the grid diagonally. The external low
density lobes are dragged away in a turbulent motion while the ISM in
the central region is essentially unaffected. Given the
particular direction of the motion, at $t=100$ Myr two 
transmitted shocks
are present, one in the leading edge of the galaxy and one in the rear 
of the galaxy (lower panel $b$). At later times, the bow shock is 
barely visible and the galaxy has a more compact aspect. 
Also in this case both the galactic and the central regions retain 
most of its ISM (Fig. 5).

Model MD-90-LO, moving face-on through the IGM, presents a cylindrical
symmetry and has been simulated with both the 3D BOH code and with
ZEUS-2D. The latter simulation used cylindrical coordinates, with
the $R$-axis and $z$-axis having the same spacing as the axes of 
3D Cartesian grid (see section 4). The two simulations show very similar
results, demonstrating the reliability of the two codes.
The model calculated with the 3D code is shown in the lower strip
of Fig. 4.

The gas flow keeps a noticeable symmetry
up to late times, when the
downstream turbulence in the tail of the flow eventually prevails. In
the upper panel $b$ the transmitted shock is clearly visible as a
wing structure. Such a structure is responsible for the ring around
the galaxy visible in the lower panel $b$. The galaxy appears
truncated on the $z=0$ plane at the radius $R_{\rm str} \simeq 5.5$
kpc, in agreement with the value predicted by equation (9) (see also
Fig. 3). No ISM is lost from the central region, and only $\sim 20\%$
of gas has been removed from the whole galaxy (Fig. 5).

This model has been run also including radiative cooling.
The morphological differences with respect to the adiabatic model
are negligible and the stripping rate differs by few percent.
 
\subsubsection{Discussion of model MD-LO}

Several galaxies are reported in literature whose distorted and
asymmetric HI distributions are claimed to be suggestive of ram
pressure stripping by the IGM. Here we consider the dIrr galaxy
Holmberg II which has structural parameters similar to those of
our model MD (Bureau \& Carignan 2002).  Actually, the appearance of
its ISM (Fig. 3 in the paper of Bureau \& Carignan) closely resembles
the gas distribution in our model MD-00-LO (lower panel $d$ in the
upper strip of
Fig. 4). This suggests that Holmberg II is moving nearly edge-on
through the IGM of the M81 group. Moreover, comparing its asymmetric
tail with our simulations, it can be concluded that the galaxy is
rotating clockwise on the sky. From a condition similar to equation
(9), together with the measurement of the galactic velocity and of
the ISM radius, Bureau \& Carignan estimate that the value of the IGM
density is $n\ga 4\times 10^{-6}$ cm$^{-3}$ at a projected distance of
475 kpc to M81. While equation (9) holds for face-on stripping,
our simple analysis shows that the model MD is not
sensitive to the value of the inclination angle $\theta$ (Fig. 3). Thus, the
evaluation of the ICM density given by Bureau \& Carignan remains
valid.

\begin{figure*}
\begin{center}
\psfig{figure=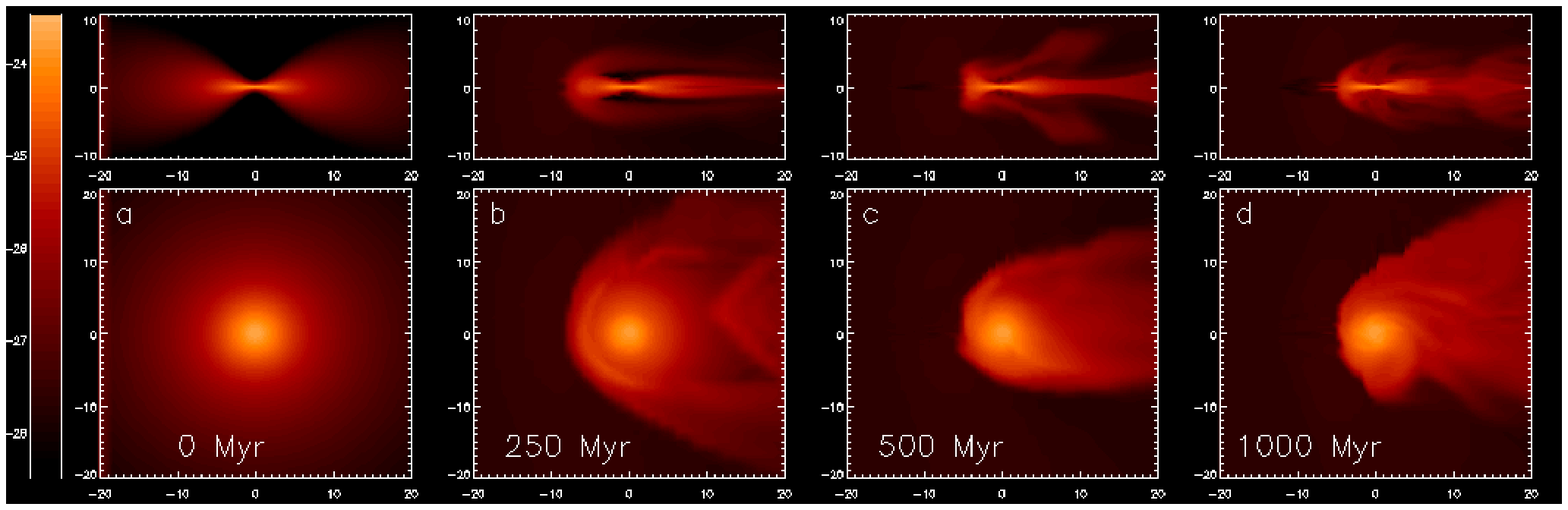,width=17.5cm,height=6.5cm}
\end{center}

\bigskip

\begin{center}
\psfig{figure=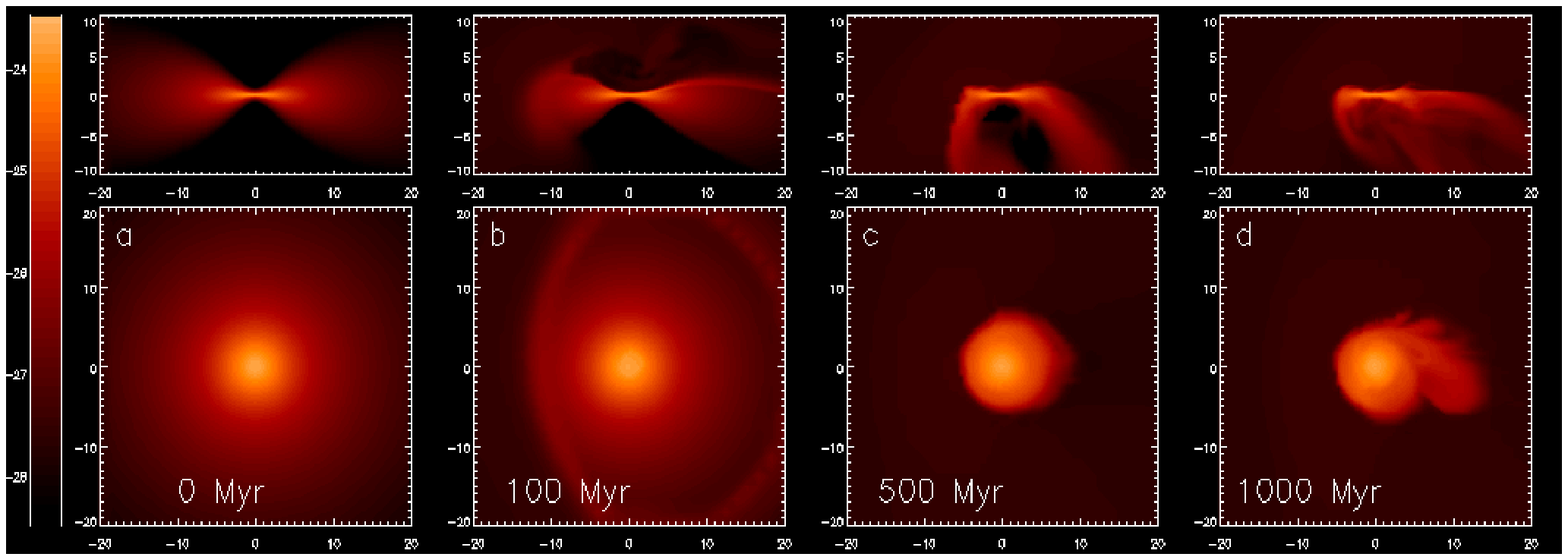,width=17.5cm,height=6.5cm}
\end{center}

\bigskip

\begin{center}
\psfig{figure=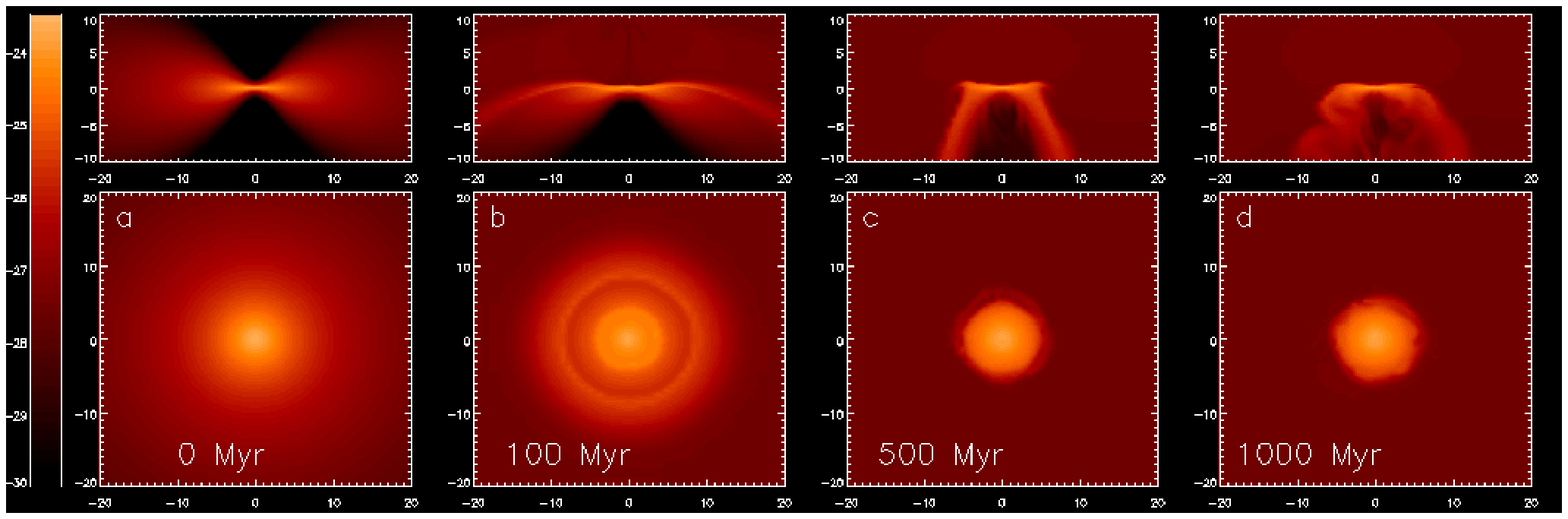,width=17.5cm,height=6.5cm}
\end{center}
\caption{Evolution of the density for model 
MD-00-LO (upper strip), MD-45-LO (middle strip) and MD-90-LO (lower strip). 
In each strip the lower panels show the density map in the
$z=0$ plane, the upper panels show the density map
in the $y=0$ plane. Axes units are kpc.}
\end{figure*}
 
Numerical simulations confirm that the results of stripping is
not very sensitive to $\theta$.
Although the effects of the interaction with the IGM in model MD
are apparent at
large radii, from Fig. 5 it is evident that
the gas in the central region is essentially unaffected
for any inclination of the galaxy. The amount of ISM in the
galactic region
is only little more sensitive to the direction of the IGM motion
(lower panel $b$). These results are consistent with the analytical
predictions given in section 3 and summarised in Fig. 3.  
The stripping radius of the edge-on model obtained in
our numerical simulation results to be a bit larger than that
predicted by the Fig. 3. This is due to the long time-scale of the
gas ablation, as discussed in section 7.

\subsubsection{Models MD-HI}

We now briefly describe the same galaxy model of the previous section,
in the case of high ram pressure.  Contrary to the previous case,
now $P_{\rm ram} = 3.2\times 10^{-12} \gta P_0 = 1.2 \times 10^{-12}$
dyn cm$^{-2}$, and the inclination angle $\theta$ becomes a crucial
parameter of the problem.
\begin{figure*}
\begin{center}
\psfig{figure=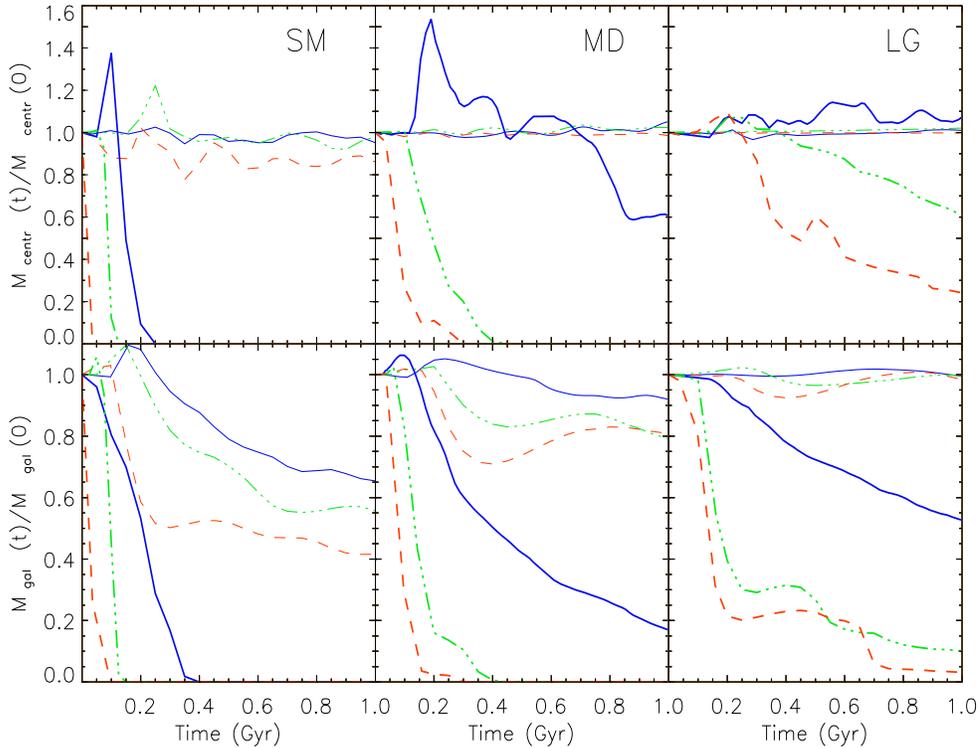,width=13cm,height=10cm}
\end{center}
\caption{Evolution of the ISM mass in the ``central region'' (upper panels) 
and in the ``galactic region'' (lower panels). The mass refers to the
cold gas with temperature $T<3 \times 10^5$ K. 
From left to right the couples of lower and upper panels refer to SM, MD and  
LG. 
Solid blue lines: edge-on models; dashed-dotted green lines: $45^{\circ}$ models;
dashed red lines: face-on models. Light and heavy lines refer to LO and HI ram 
pressure models, respectively.}
\end{figure*}

In Fig. 5 it
is evident that, for the models MD-90-HI and MD-45-HI,
the gas is completely removed from both the galactic and central regions
after $\sim 2-4 \times 10^8$ yr.
The edge-on model, instead, is more resistant to the stripping
and retains more than $60\%$ of its
initial ISM in the central region after $t=1$ Gyr.  
The gas amount in the whole galaxy has been reduced to $\sim 20 \%$
of the initial value.
This result appears contrary to
the simple theoretical expectation of complete gas removal given by Fig. 3.
However, $\tau_{\rm str} > 1$ Gyr and 
$M_{\rm gal}$ is indeed still decreasing at $t=1$ Gyr,
as it is clear from Fig. 5.
We verified that all the gas is removed from the galaxy
after $t\sim 1.7$ Gyr.

As can be seen in Fig. 5, $M_{\rm centr}$
increases up to a maximum at 200 Myr for model MD-00-HI, and
oscillates before decreasing by a factor of $\sim 2$. At this point
the radius of the galaxy has been reduced to $R_{\rm gal}\sim 3$ kpc. The
first maximum is related to the passage of the transmitted shock
through the galaxy, while the oscillations reflect fluctuations
of the ISM on the characteristic time scale ($\sim 2 \times 10^8$ Myr)
of the sound crossing time of the central region. The magnitude of
such oscillations strongly depends on our definition of central
region; on larger scales the mass stripping proceeds in a more uniform
fashion, as shown by the smooth behaviour of $M_{\rm gal}$.

Given the different geometry of the IGM impact, the
$\theta=45^{\circ}$ and face-on models are more exposed to the effect
of the ram pressure,
and the stripping is complete after $t\sim 3-4 \times 10^8$ yr
(Fig. 5).

\subsection{Other models}
\subsubsection{Models SM}

Now we examine the case of the smaller galaxy. Because of its shallower
potential, it is the most sensitive to the influence of the IGM. From
Fig. 3 we expect that the ISM evolution is not
strongly dependent on
the inclination angle.

The simulations broadly confirm this expectation. For model SM-LO, the central 
region is basically unaffected by the ram pressure for every $\theta$.
On the larger scale, $M_{\rm gal}$ rapidly decreases by a factor of 
$\sim 2$ in the face-on case because of the instantaneous stripping.
After $t\sim 300$ Myr the stripping rate greatly reduces and $M_{\rm gal}$
decreases very slowly.
The edge-on model, instead, loses mass at a lower and more uniform rate.
This is because the continuous stripping is more effective than the 
instantaneous stripping, as expected given this specific geometry.
Again, $M_{\rm centr}$ is essentially unaffected being $P_0 > P_{\rm ram}$.
At the end of the
simulation the galactic radius is $\sim 2$ kpc and the ISM has
a quite compact aspect.

In the model with stronger ram pressure (SM-HI), instead, the ISM is
completely removed from the central region in few $10^8$ yr and
leaves the galaxy in $3-4 \times 10^8$ yr for any
inclination.  This is what we exactly expect from the
theoretical analysis, so this kind of galaxy would show only its stellar
component with no or very little gas content (see section 6).
For model SM-00-HI the gas in the galactic region 
temporary increases before to drop rather quickly. The density increase 
is due to the same mechanism discussed in the previous section for 
model MD-00-HI.

\subsubsection{Models LG}

This galaxy with its stronger gravitational field and central density is, 
of course, the least affected by ram pressure stripping.
For the models with lower ram pressure (LG-LO) 
the stripping is minimal (see Fig. 5).
In both regions under analysis the ISM mass does
not vary significantly, for all values of $\theta$.
The final radius is quite 
large ($>20$ kpc) and this galaxy is rather unperturbed also at large radii.

\begin{figure*}
\begin{center}
\psfig{figure=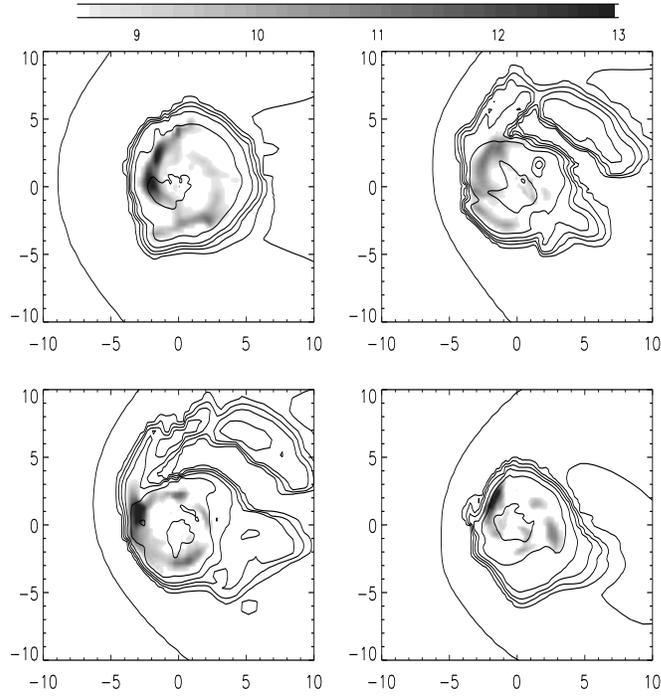,width=9cm,height=9cm}
\end{center}
\caption{ISM face-on column density maps for model LG-45-HI. Only values
larger than the critical value
$\Sigma_{\rm g}=9 M_{\odot}$ pc$^{-2}$ are displayed. The time sequence is
$t=(250, 350, 400, 550)$ Myr starting from the left-top 
panel and moving clockwise.
The grey scale indicates $\Sigma$ in units of
$M_{\odot}$ pc$^{-2}$.  Contours represent the logarithm of ISM
density on the galactic plane $z=0$. The contour values are
-26.5, -26.0, -25.5, -25.0, -24.5, -24.0 and -23.5. Axes units are kpc.} 
\end{figure*}

\begin{figure*}
\begin{center}
\psfig{figure=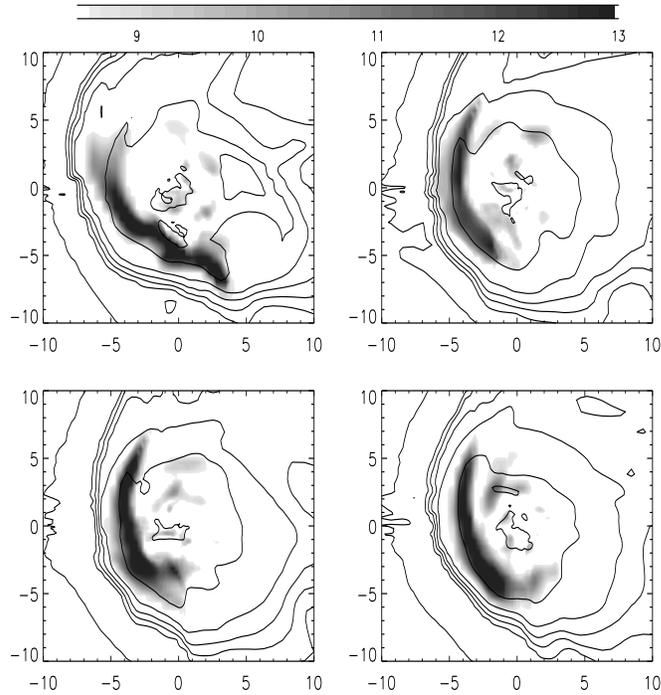,width=9cm,height=9cm}
\end{center}
\caption{Same as Fig. 6 for model LG-00-HI.
The various panels refer to $t=(380,470,510,560)$ Myr.
The contour values for the gas density are again
-26.5, -26.0, -25.5, -25.0, -24.5, -24.0 and -23.5.}
\end{figure*}

Even the higher ram pressure in models LG-HI is lower 
than (but close to) the central ISM pressure,
$P_{\rm ram}/P_0 \sim 0.66$.
As discussed above, we expect that for $P_{\rm ram} \sim P_0$
the inclination angle of the
galaxy plays a crucial role. Indeed, as shown in Fig. 5, in the edge-on case 
$M_{\rm centr}$ is essentially unaffected by the ram pressure, 
while it decreases
for the $\theta=45^{\circ}$ model and even more for the face-on model.
$M_{\rm gal}$ shows a behaviour similar to that of model SM-LO.
In the edge-on case the mass loss is rather steady, while for models
LG-45-HI and LG-90-HI the action of both the instantaneous and continuous
stripping are apparent.

\subsubsection{Models NR}

We describe here the evolution of the ISM in a non rotating galaxy.
As discussed in section 2.2, we are particularly interested to the fate
of the ISM located at large radii. A spheroidal, extended gaseous
halo could suffocate the galactic winds triggered by a central
starburst, with important consequences for the chemical evolution of
the galaxy.
The ISM in this model, which has the same gravitational potential
of model MD, but a higher ISM temperature (see section 2.3),
extends initially to $R_{\rm ISM} \sim z_{\rm ISM} \sim 4$ kpc.
The gas density distribution is rather flat near the galactic plane
because the influence of the stellar disc gravity, and becomes
nearly spherical for $R,\, z \gta 2$ kpc. The initial ISM surface
density peaks at the center, $\Sigma (0) \sim 11$ M$_\odot$ pc$^{-2}$,
and drops rapidly with $R$, $\Sigma \sim 1.4$ M$_\odot$ pc$^{-2}$
at $R\sim 1$ kpc and $\Sigma \sim 0.05$ M$_\odot$ pc$^{-2}$
at $R\sim 2$ kpc.
The central and galactic regions for this models are defined as
two spherical volumes defined by $r \leq 1$ and
$r \leq 4$ kpc respectively. The initial gas masses in these
regions are $M_{\rm centr} = 1.3 \times 10^7$ M$_\odot$
and $M_{\rm gal} = 1.7 \times 10^7$ M$_\odot$ respectively.
We report here only on face-on NR models (NR-LO and NR-HI).
More non rotating galaxies will be discussed elsewhere
(Marcolini et al. in preparation).

Model NR-LO loses gas at a nearly constant rate 
$\dot M_{\rm gal} \sim 2.1 \times 10^6$ \msun Gyr$^{-1}$
and $\dot M_{\rm centr} \sim 1.1 \times 10^6$ \msun Gyr$^{-1}$.
For pedagogical reasons we run this model for a very long time
$t=10$ Gyr. 
The continuous ablation of the outer gas layers is the
main process at work in this model, as expected being 
$P_0 > P_{\rm ram}$. At $t\sim 300$ Myr the gas at large
radii has been already stripped away and the ISM has a compact
aspect. At this time, the ISM extends to a radius $R\sim 1.4 $ kpc
on the equatorial plane $z=0$.
Hereafter the stripping slowly reduces the size of the ISM
which is completely removed in $t\sim 10$ Gyr (interestingly,
being $P_0 > P_{\rm ram}$, an
incomplete stripping is expected according to the simple theory discussed
in section 3).
For most of the time 
the ISM distribution is asymmetric with respect to the $z=0$ plane,
extending to $z\sim 0.5$ kpc upstream and to $z \sim -1 \;\div \; -1.8$ kpc
in the cometary tail, depending on the time considered.
We conclude that extended, spheroidal gaseous halos 
are unlikely to survive
under the effect of (moderate) ram pressure stripping. 

A backflow is generally present in this model, but it scarcely
influences our results. Balsara, Livio and O'Dea 1994, Stevens, 
Acreman and Ponman 1999, Mori \& Burkert 2000, among others,
studied this backflow in some detail. In their calculations
the accretion inflow occurs quasi-periodically, producing an 
oscillation in the rate of gas stripping.
The accretion inflow disappears when the gas is not
allowed to cool in the models of Balsara, Livio and O'Dea (1994), but
it is instead present in the simulations of Mori \& Burkert (2000) 
who neglect
radiative losses. As discussed in section 6, the parameters of our 
models imply a general unimportance of the accretion backflow, independently
on the effect of the radiative cooling.

In model NR-HI all the ISM is removed in $\sim 500$ Myr, as expected
from the relation $P_0 < P_{\rm ram}$. This time is in reasonable
agreement with the timescale $\tau_{\rm str}$ given by equation (12),
$\tau_{\rm str} \sim 400$ Myr, with 
$\dot M \sim \pi R_{\rm gal}^2 \rho_{\rm IGM} v_{\rm IGM} \approx
2.5 \times 10^{23}$ g s$^{-1}$, using an average 
$R_{\rm gal} \sim 1$ kpc.

\section{discussion and conclusions}

We run several 3D and 2D hydrodynamic simulations of ram pressure stripping
in galaxies moving through an external medium. 
We focused on the poorly investigated problem of disky dwarf galaxies
located in small groups, where they are in fact more frequently observed.
The ram pressures considered, $P_{\rm ram} = 8\times 10^{-14} - 
3.2 \times 10^{-12}$ dyn cm$^{-2}$ are one or two orders of magnitude lower
than that expected in rich clusters, the subject of most previous studies.
We run a set of models varying galactic mass (together with
the relative size and gas content), inclination angle $\theta$ between 
galactic plane and orbital plane, and ram pressure strength.

Following Gunn \& Gott (1972), we made simple analytic previsions of
the disk radius beyond which the ISM is completely stripped in face-on
models (section 3). We also extended the condition for ram pressure
stripping to the edge-on models.  Compared to the analytic estimates,
the radii of our edge-on models tend to be a little larger after 1 Gyr.
We can consider the time-scale $\tau_{\rm str}$ given in equation (12)
a rough estimate of the time needed by the ram pressure to ablate the
galaxy to its final radius. In our models $\tau_{\rm str}$ may be as
long as a few Gyr, as confirmed by our numerical simulations.  
In fact, in some model the stripping is still going on at $t=1$ Gyr,
and thus the final radius is not yet
reached at the end of the run.
Given the simple assumptions on
which the analytical model is based, the agreement with the numerical
simulations is quite satisfactory.

We find that the inclination angle often plays a minor role. For $P_{\rm
ram}>P_0$ the galaxy is stripped for any inclination. In the case of
$P_{\rm ram}<P_0$ the amount of mass lost is rather independent of
$\theta$, with the edge-on models showing a slight
tendency to retain more ISM (see Fig. 5). 
Only for $P_{\rm ram}\sim P_0$ the ability
of the galaxy to retain its ISM is significantly 
larger for $\theta =0^{\circ}$.
From the analysis presented in section 3 the final ISM radius $R_{\rm str}$
is expected to be more sensitive to $\theta$ for larger galaxies,
as it is apparent comparing the continuous curves (face-on galaxies) 
to their correspondent dashed curves in Fig. 3: only for the most 
massive galaxy the two curves differ significantly.
The numerical simulations, however, show a stronger dependency
of the effectiveness of the stripping on $\theta$ (and time).
For both models MD and LG the mass of gas retained by the galaxy 
is indeed a sensitive function of $\theta$ for the stronger 
ram pressure (Fig. 5).

We note that when $P_{\rm ram}<P_0$, the ISM can still be completely stripped,
as discussed in section 5.2.3 for a non-rotating, nearly spherical galaxy
(model NR-L0).
However, in this case
the time scale for the complete gas removal is very long, 
$\sim 10$ Gyr. This may imply that ancient dwarf galaxies in old poor
groups, where the ram pressure is weak, may have their ISM removed. 
We note that the galaxy population in X-ray bright groups is indeed
dominated by early-type (gas poor) dwarfs (e.g. Tran et al. 2001).

There is some debate in literature on the successors of stripped
dwarf irregular galaxies. Conselice, Gallagher, \& Wyse (2001) argue
that dwarf elliptical galaxies presently seen in the core of the Virgo
cluster cannot
originate from field dwarfs accreted by the cluster. 
If this would be the case, very few post-stripped dwarf irregulars
would be present in Virgo (Lee, McCall, \& Richer 2003).
However, other lines of evidence indicate that ram pressure stripping
could have converted many dIrrs in the present-day dEs 
(Lee, McCall, \& Richer 2003; Conselice et al. 2003).
In this scenario, tidal interactions with massive galaxies
may provide the tool to transform rotationally supported stellar disks
in slowly rotating, spheroidal galaxies
(Mayer et al. 2001). 

As discussed in section 5.2.3, several authors report the occurrence of
an accretion inflow into the galactic core from the downstream side
due to the gravitational force (Balsara, Livio and O'Dea 1994,
Stevens, Acreman and Ponman 1999, Mori \& Burkert 2000).  This flow
influences the stripping rate and may cause the appearance of a second
bow shock (Balsara et al. 1994). 
We do not find any of the complexities described above in
our rotating models. This is due to physical reasons and cannot be
ascribed to the lower spatial resolution of our 3D simulations
compared to the resolution of the 2D models described in the
aforementioned papers.  In fact, the behaviour of flows past
gravitating bodies is in many cases determined by the ratio
$\xi={2GM \over
R(v^2+c^2)}$ (Balsara et al. 1994), 
where $M$ and $R$ are the mass and the radius of the
gravitating body, $v$ is the relative velocity and $c$ the sound
speed. In our models $\xi < 1$, and thus accretion is not supposed to
occur. A (variable) backflow is present is some models, 
especially in model NR,
but it does not accrete onto the galaxy and it does not influence the
stripping rate 

The efficiency of star formation is a strong function of the ISM
pressure (Elmergreen \& Efremov 1997). As the transmitted shock
propagates through the galaxy, it may promote the collapse of
molecular clouds inducing star formation. Obviously, we can not follow
this phenomenon in our one-phase ISM models (see below). However, from the
temporary increase of the ISM content in the galactic region of some
models (see Fig. 5), we suspect that favourable conditions for star
formation are possibly achieved in galaxies undergoing ram pressure
stripping. More accurate
considerations may be done taking into account the evolution of the
ISM surface density $\Sigma_{\rm ISM}$ on the galactic plane of our
models.  Actually, episodes of star formation are believed to take
place in regions where the ISM surface density exceeds the critical
value $\Sigma_{\rm cr} \sim 9$ $M_{\odot}$ pc$^{-2}$ (Gallagher \&
Hunter 1984; Skillman 1987, 1996), although this is not a sufficient
condition (Dohm-Palmer et al. 2002).  As an example, Fig. 6  shows
maps of $\Sigma_{\rm ISM}$ on the galactic plane of the model LG-45-HI
at several times. $\Sigma_{\rm ISM}$ is
computed integrating in the interval $-z_{\rm centr}<z<z_{\rm centr}$ to
neglect spurious contributions from clouds of gas located at
larger $|z|$.  Only values $\Sigma_{\rm ISM}>\Sigma_{\rm cr}$ are
shown in Fig. 6, indicating regions where star formation is likely
to occur.  These regions are distributed rather stochastically both in
space and time on a time span of 600 Myr. A stochastic pattern
of star-forming regions is what Dohm-Palmer et al. (2002) find in the
dwarf galaxy Sextans A. Of course,
other external causes, such as tidal interactions with nearby
galaxies, may also be responsible of star formation episodes. Moreover, in
our models the largest supercritical values of $\Sigma_{\rm ISM}$
never occur in the galactic centre, where instead strong star burst
are observed. We showed, however, that the IGM ram pressure
in small groups may create the conditions for moderate and long
lasting star formation in dwarf galaxies.

In general, the surface density increases more easily for small values
of $\theta$ because a part of the gas is pushed to smaller galactic
radii in the beginning of the stripping. This effect is also found by
Vollmer et al. (2001), and Hidaka \& Sofue (2002) find an increment of
the molecular gas fraction in the central region of spiral
galaxies suffering mild stripping. Possible episodes of star formation
are expected to be triggered in this cases. In fact, our models
LG-00-HI and MD-00-HI, which temporary increase their galactic gas
content (Fig. 5), also present regions of supercritical surface
density (Fig. 7). 
However, no direct link exists between mass content increase and surface
density increase. In fact the model SM-00-HI, which also shows a
``spike'' in its galactic ISM content, never reaches a gas surface
density larger than $\Sigma_{\rm cr}$, and is completely stripped
without any previous episode of star formation. On the contrary, in
the models MD-45-HI and LG-45-HI, which never increase their galactic
gas content, $\Sigma_{\rm g}$ episodically exceeds the critical value
close to the galactic centre.  Finally, the lowest value of ram
pressure (LO) can not compress the ISM enough to obtain $\Sigma_{\rm
g}>\Sigma_{\rm cr}$ for any value of $\theta$ and of galactic mass.
From the above discussion we conclude that star bursts in dIrrs can be
triggered by ram pressure in poor groups, but the galaxy must cross
the group core, where larger values of $P_{\rm ram}$ occur.

All the above conclusions are obtained in the assumption that the ISM
is distributed in a smooth fashion. Actually real ISM is a complex
gaseous medium in which HI and massive molecular clouds are present.
HI clouds have relatively low densities ($\sim 1$ cm$^{-3}$) and large
sizes ($\sim 0.5$ kpc) and readily react to external perturbations
given by $P_{\rm ram}$. We thus expect that their cumulative behaviour
can be correctly described in the frame of single-phase ISM
distribution.  As pointed out above, we neglect the presence of
molecular clouds because they are too small to be resolved by our
simulations. In any case, given their large densities (100 cm$^{-3}$)
and small radii ($\sim 10 $ pc) they are much less affected by
stripping processes (e.g. Raga et all. 1998). Differences in the
response of HI and molecular clouds to ram pressure are found by Sofue
(1994a,b).

In a successive paper
we will consider the effect of a star burst on the chemical and dynamical
evolution of the ISM of dwarf galaxies undergoing ram pressure stripping.

\section*{Acknowledgements} 
We are grateful to the referee for his/her helpful suggestions which
improved the presentation of the paper.
We acknowledge financial support from National Institute for
Astrophysics (INAF).
The simulations were run at the CINECA Supercomputing Centre with CPU time
provided by a grant of the National institute for Astrophysics (INAF).

{}

\label{lastpage} 
\end{document}